\documentclass[draft]{agujournal2019}
\usepackage{url} 
\usepackage{soul}
\usepackage{amssymb,amsmath}
\draftfalse
\journalname{Water Resources Research}
\begin{document}

%
%

\title{Practical data-driven flood forecasting based on dynamical systems theory: Case studies from Japan}
%
%



\authors{Shunya Okuno\affil{1,2}, Koji Ikeuchi\affil{3}, and Kazuyuki Aihara\affil{1,4}}

\affiliation{1}{Institute of Industrial Science, The University of Tokyo}
\affiliation{2}{Weather Disaster Prevention Section, Kozo Keikaku Engineering Inc.}
\affiliation{3}{Department of Civil Engineering, School of Engineering, The
    University of Tokyo}
\affiliation{4}{International Research Center for Neurointelligence (WPI-IRCN), The University of Tokyo}

\correspondingauthor{Shunya Okuno}{okuno@sat.t.u-tokyo.ac.jp}



\begin{keypoints}
    \item We develop a data-driven method to forecast river stages based on dynamical systems theory
    \item It can handle unexperienced water levels and make forecasts using fewer past water rise events
    \item The method outperforms existing methods for actual rivers, including a physical runoff model
\end{keypoints}

%

%
%

\begin{abstract}
    Data-driven flood forecasting methods are useful, especially for the rivers that lack hydrological information to build physical models.
    Although these former methods can forecast river stages using only past water levels and rainfall data, they cannot handle previously unexperienced water levels easily, and require a large amount of data to build accurate models.
    Here, we focus on phase-space reconstruction approaches, and develop a practical data-driven forecasting method to overcome the existing problems.
    The proposed method can handle the unexperienced water levels and provide forecasts using only a small number of water rise events.
    We apply the method to data from actual rivers, and it achieved the best forecast performance among existing methods, including a physical runoff model, a data-driven multi-layer perceptron, and a conventional method based on phase-space reconstruction.
    In addition, the proposed method also forecasted the exceedance of the evacuation warning level 6 h earlier for steep rivers.
    Given its performance and maintainability, the proposed method can be applied to many actual rivers for early evacuation.
\end{abstract}

\section*{Plain Language Summary}
Decision makers especially find it useful to forecast river stages when deciding to evacuate citizens and minimize flooding damage.
The most common forecasting methods are based on physical models, which usually require many kinds of hydrological details.
Some approaches are data-driven and provide forecasts from time series of past river stages and rainfall. However, they require large amounts of data for accurate forecasting, and it is difficult to forecast previously unexperienced flood magnitudes.
Considering the recent progress in nonlinear science, we present a data-driven approach based on the dynamical systems theory to overcome these existing problems.
Given its accuracy and maintainability, the proposed method can be applied to many rivers and is useful for deciding whether to evacuate in flooding situations.

%


%
%
%
%
\section{Introduction}
The severity of climatic natural disasters has increased over recent years,
and repetitive flooding incidents have led to the loss of numerous lives.
In Japan particularly, rapid flooding of small and steep rivers has posed a serious problem.
For example, in the year 2016, Omoto River flooded in a few hours, leading to the loss of many lives, and heavy rainfall in northern Kyushu in 2017 killed many people because of the resulting rapid floods in the associated small rivers.
Although inexpensive water gauges are being used to monitor water levels of small rivers in recent years, it is difficult to take timely and proper evacuation actions using information on the water level at a particular point in time alone, because the water rise tends to be rapid.

Flood forecasting systems help decision makers ensure proper broadcasting of evacuation notices and instructions; however, most small rivers lack such systems.
Flood forecasting is generally conducted by applying runoff analysis, which requires many kinds of data, including the water level, precipitation, discharge, catchment characteristics, and river morphology (such as channel cross-sectional data).
In the case of small rivers, it is difficult to accurately determine the channel morphology and measure the discharge.
Furthermore, considering the costs required for adjusting model parameters, runoff analysis methods are not always appropriate to model such small rivers.

To forecast the water level without detailed-hydrological information, we focus on data-driven approaches---which require only water level and precipitation data---because of the reasons described above.
One of the major approaches is to assume functions and fit their parameters using past observations; e.g., as accomplished by conventional auto regressive (AR) models and artificial neural networks.
The performance of AR models is not satisfactory because they do not consider nonlinearity of the runoff process. Thus, some studies have applied deep neural networks to improve performance \cite{liu2017, hitokoto2017}.
Deep neural networks show good performance, as shown in existing studies. However, two main issues persist.
The first relates to performance when considering previously unexperienced flood magnitudes.
As neural network models assume black box functions and perform interpolations using past data with these functions, they may show poor performance for data that exceed the range of the training data \cite{hitokoto2017},
and it is difficult to analyze the behavior for such data.
The other issue concerns the amount of data required.
In general, deep neural networks achieve outstanding performance with regard to adjusting numerous parameters through learning with a huge amount of data, but it is difficult to obtain such data on many water rise events.
In fact, \citeA{hitokoto2017} employed shallow and two hidden layers for this purpose.

In this study, we use a model-free approach based on the dynamical systems theory.
This approach assumes that the runoff system is a deterministic dynamical system,
and the approach reconstructs a possible attractor of the original system only from observed time series on the basis of embedding theorems \cite{Takens1981, Sauer1991a}.
This approach has been discussed and widely applied to runoff analyses by many researchers such as \citeA{Jayawardena1994}, \citeA{Porporato1997}, \citeA{Liu1998}, \citeA{Laio2003}, and \citeA{Costa2012}.
However, they simply applied conventional nearest neighbor approaches without considering recent progress in nonlinear forecasting techniques.
In recent years, several outstanding forecasting approaches with multiple embeddings have been proposed in the field of nonlinear science, especially for short- and high-dimensional data \cite{Ye2016, Ma2018a, Okuno2019, Okuno2019a}.
These methods are considered to be suitable for river stage forecasting, especially for small rivers, which usually contain multidimensional but limited samples.
On the other hands, we need to modify the forecast map to consider the previously unexperienced magnitude of floods when we apply the recent methods to actual rivers.

Herein, we propose a practical river stage forecasting technique based on the dynamical systems theory.
We first develop a forecast map to treat previously unexperienced river water levels by combining existing recent forecasting techniques.
Then, we apply the proposed method to actual rivers, and evaluate its validity for practical use.
In particular, we evaluate not only the accuracy of the proposed approach by comparing the results with those of existing methods, but we also assess the lead time for evacuation, behavior of the approach for unexperienced river stages, and the number of flood events required to forecast evacuation.

\section{Materials and Methods}
In this section, we first summarize the prerequisites of the dynamical systems theory.
Then, we explain the scheme to be applied to the river stage while considering rainfall forecast, and propose a method to treat previously unexperienced river stages in an actual system.

\subsection{Attractor Reconstruction by Delay Embedding}
According to embedding theorems \cite{Takens1981,Sauer1991a} in the dynamical systems theory, we can reconstruct the attractor of the original dynamical system only from observed time series.
Suppose that we have the dynamical system $f:\mathbb{R}^m \to \mathbb{R}^m$ and the observation function $g:\mathbb{R}^m \to \mathbb{R}^n$, as follows:
\begin{eqnarray} \label{eq:system}
    \frac{\mathrm{d}x(t)}{\mathrm{d}t}=f(x(t)), \\
    y(t)=g(x(t)),
\end{eqnarray}
where $x(t) \in \mathbb{R}^m$ is the state of the dynamical system, and $y(t) \in \mathbb{R}^n$ is the observed vector.
Although it is impossible to observe all variables of $x(t)$ for most cases, we can reconstruct the original attractor using delay coordinates $v(t) \in \mathbb{R}^E$ according to the embedding theorems \cite{Takens1981,Sauer1991a}. For details, see \ref{sec:embedding}.
These embedding theorems ensure that the map from the original attractor to the reconstructed one has a one-to-one correspondence with an appropriate $E$.
We can analyze the observed time series using the reconstructed attractor. This is the basic principle of nonlinear time series analysis.

\subsection{Reconstruction of Runoff Process from Observed Time Series}
We can reconstruct the rainfall runoff process by delay embedding, as shown in existing studies \cite{Jayawardena1994, Porporato1997, Liu1998, Laio2003, Costa2012}.
Here, we explain the iterative procedure to apply delay embedding to the runoff process.
For instance, suppose we observe river stage $y_{s}(t)$ at site $s$, and rainfall data $y_r(t)$ at site $r$.
We consider the following delay coordinates to treat the data as multivariate time series for example:
\begin{equation}
    v(t) = [y_s(t),y_s(t-1),y_r(t),y_r(t-3)].
\end{equation}
When we have a map $\psi$ such that $v(t+1) = \psi(v(t))$, we can obtain the forecast value at time $t+1$ as $\hat{y_s}(t+1)$.
Simultaneously, we can create $\hat{v}(t+1)$ as follows:
\begin{equation}
    \hat{v}(t+1) = [\hat{y}_s(t+1),y_s(t),\hat{y}_r(t+1),y_r(t-2)].
\end{equation}
We can forecast the desired time step to iterate this procedure.
If rainfall forecast $\acute{y}_r(t+1)$ is available, we can substitute $\hat{y}_r(t+1)$ with $\acute{y}_r(t+1)$.
The rainfall forecast is crucial, especially for long-term forecasts for steep rivers.
In this paper, we used the rainfall forecast for Japanese rivers because these rivers are mostly steep, and it is easy to obtain the rainfall forecast for a desired point in Japan.
Note that we considered actual rainfall as forecast for these cases because we discuss the error of the proposed methods, not the error of the rainfall forecast.
Regarding the competition data described in \ref{sec:annex}, we did not use the rainfall forecast.

Although we described the case of two variables (one is the river stage and the other is the rainfall) here,
we can consider multiple rainfall sites and river stages, including upstream rivers, for actual applications.
See equation~(\ref{eq:delay_embedding}) in \ref{sec:embedding} for the general multivariate delay coordinates.
In general, the number of possible embeddings grows combinatorially with the number of variables, that is, the number of river stages and rainfall sites.
Several existing studies have exploited this property \cite{Ye2016, Ma2018a, Okuno2019a}.
These methods yielded single forecasts by combining multiple forecasts based on multiple embeddings.
We employed the method of \citeA{Okuno2019a}, who solved an optimization problem by minimizing the forecast error (see \ref{sec:ga}).

\subsection{Forecast map to treat unexperienced river stages}
Here, we propose a map $\psi$ to forecast an unexperienced magnitude of the water level.
The method of analogues \cite{Lorenz1969} is a conventional forecasting method based on attractor trajectories.
The method first searches the set of nearest neighbors $\mathcal{I}(t)$ of the current query $v(t)$ from $\{v(t') \mid t' \in \mathcal{T}_{train}\}$, where $\mathcal{T}_{train}$ is the set of time indices of the training data.
Then, the method forecasts $v(t+1)$ using the forward trajectory path of $\mathcal{I}(t)$.
In this paper, we propose a forecasting method that considers the correction term ${\rm diag}(\lambda)z(t)$, as shown in equation~(\ref{eq:forecast}):
\begin{eqnarray}
    z(t) := v(t) - \sum_{t' \in \mathcal{I}(t)}w(t')v(t'),  \\
    \hat{v}(t+1) = \sum_{t' \in \mathcal{I}(t)} w(t')v(t'+1) + {\rm diag}(\lambda)z(t), \label{eq:forecast}
\end{eqnarray}
where $\lambda \in \mathbb{R}^E$.
The corresponding weight $w(t')$ for $\mathcal{I}(t)$ is optimized based on \citeA{Hirata2014}:
\begin{eqnarray} \label{eq:opt_weight}
    {\rm minimize}_{w} \sum_{t' \in \mathcal{I}(t)} \| z(t') \|, \\
    s.t. \sum_{t' \in \mathcal{I}(t)} w(t') = 1, \\
    w(t') \geq 0 \ \forall t' \in \mathcal{I}(t),
\end{eqnarray}
where $\|\cdot\|$ is an appropriate norm.
Although \citeA{Hirata2014} formulated the linear programming problem corresponding to the $L^\infty$ norm, we can select appropriate norms.
As we apply the Euclidean distance for the neighboring search in this study, we employ the $L^2$ norm.
With this assumption, we can quickly obtain the global optimal solution of equation~(\ref{eq:opt_weight}) because the problem is a convex quadratic programming problem.

\begin{figure}[tb]
    \centering
    \includegraphics[width=.7\linewidth]{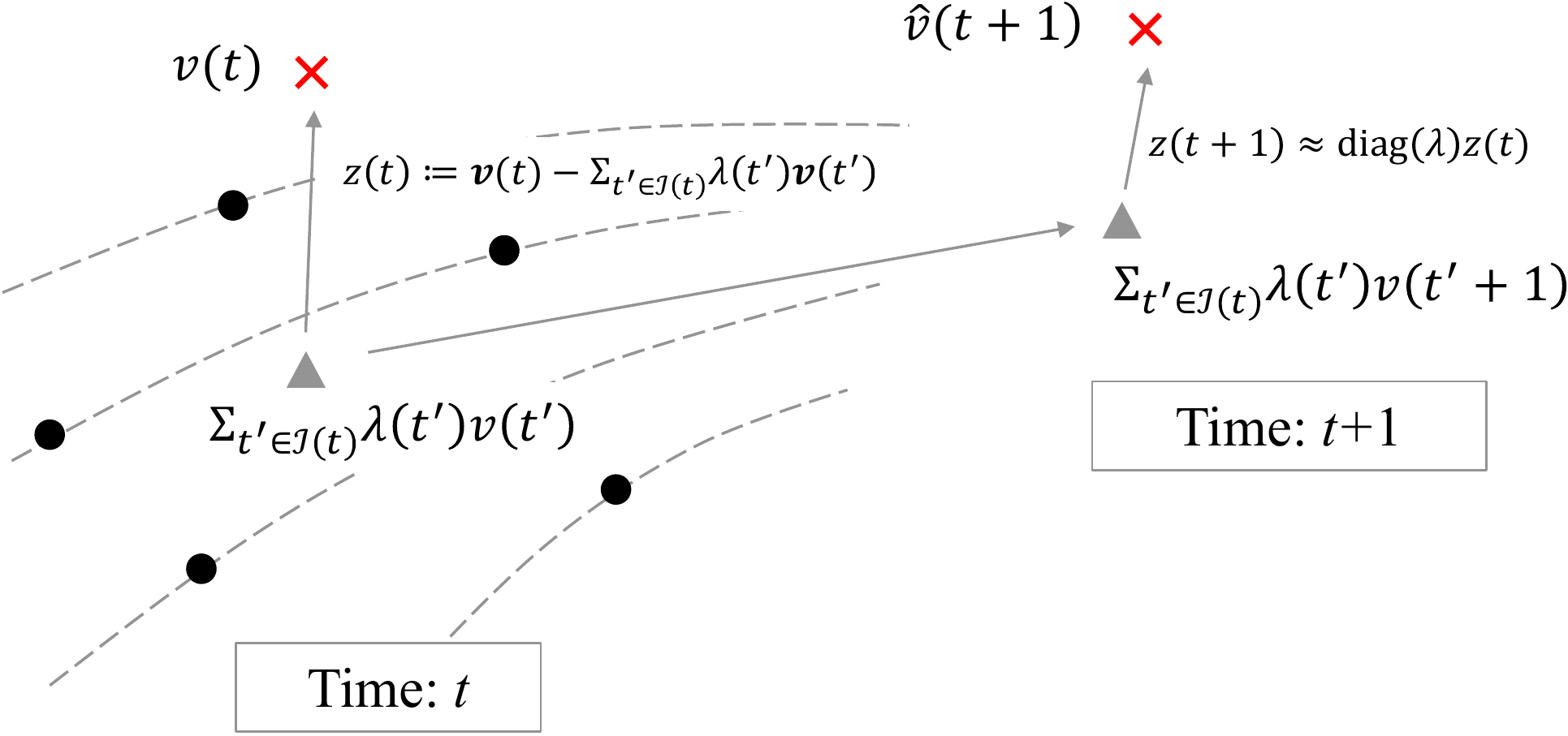}
    \caption{Schematic of the proposed forecasting map. The black dots represent the neighboring points of the current query $v(t)$ and the dashed lines represent the corresponding trajectories in reconstructed state space. The gray triangles denote the barycenters of the neighboring points at time $t$ and $t+1$, and the red cross marks refer to the current queries and corresponding forecasts, respectively. The proposed map corrects $z\left(t+1\right)$, which corresponds to the difference between the forecast $v\left(t+1\right)$ and the barycenter of the neighboring points at $t+1$.}
    \label{fig:method_of_analogues}
\end{figure}

The term ${\rm diag}(\lambda)z(t)$ corrects $z(t+1)$, which corresponds to the difference between $v(t+1)$ and the barycenter of the neighboring points at $t+1$.
The schematic of the correction term appears in Figure~\ref{fig:method_of_analogues}.
For instance, when the identity matrix is given for ${\rm diag}(\lambda)$,
we can forecast $t+1$ by simply offsetting $z(t)$ even if $v(t)$ includes an unexperienced river level that is not included in the training data.
In this study, we estimate ${\rm diag}(\lambda)$ to consider the change in $z(t)$ as $z_i(t+1) \approx \lambda_i z_i(t)$ for each $i$ as follows:
\begin{equation}
    {\rm minimize}_{\lambda_i} \sum_{t' \in \mathcal{I}(t)} \left[z_i(t'+1) - \lambda_i z_i(t') \right]^2.
\end{equation}
The solution is given by
\begin{equation}
    \lambda_i = \left[\sum_{t' \in \mathcal{I}(t)} v_i(t')v_i(t'+1) \right] / \left[\sum_{t' \in \mathcal{I}(t)} v_i(t')^2 \right].
\end{equation}
The correction term enables the forecasting of unexperienced river stages to some extent.
Note that $\lambda_i$ is approximately one for most cases because $z(t)$ does not change radically by a single time step.
If the estimation of $\lambda_i$ is unstable, we can set $\lambda_i$ to lower and upper limits.

We combine multiple forecasts based on the proposed map using the multiple embedding framework \cite{Okuno2019a}.
On application of the framework, we can obtain accurate and stable forecasts with a small number of samples even if each forecast is not very accurate.

\section{Results}
\subsection{Hiwatashi Gauging Station for the Oyodo River System}
We compared the proposed method with an existing study \cite{hitokoto2017}, which assessed a distributed runoff model, a deep neural network (multi-layer perceptron), and a hybrid model of the two.
We also compared the proposed method with a conventional local linear method \cite{Farmer1987} based on delay embedding.
We forecasted the river stages at the Hiwatashi gauging station on the Oyodo river system, whose rainfall catchment area is $861\textrm{km}^2$ wide.
We fetched hourly precipitation series of 14 precipitation stations and hourly river stage series of 5 gauging stations from 1990 through 2014 on the website of the Water Information System \cite{mizumizuweben}.
Following \citeA{hitokoto2017}, we extracted 24 sets of water rise events that exceeded 6.0 m.
We treated actual precipitation as its forecast to evaluate the error of the water level forecast without any rainfall forecast error.
We forecasted up to 6 h ahead every hour and tested the performance with the top four maximum water rise events.
For details about the data, see \citeA{hitokoto2017}.
Note that we did not employ any feature extraction unlike the existing study.

\begin{figure}
    \includegraphics[width=1.\linewidth]{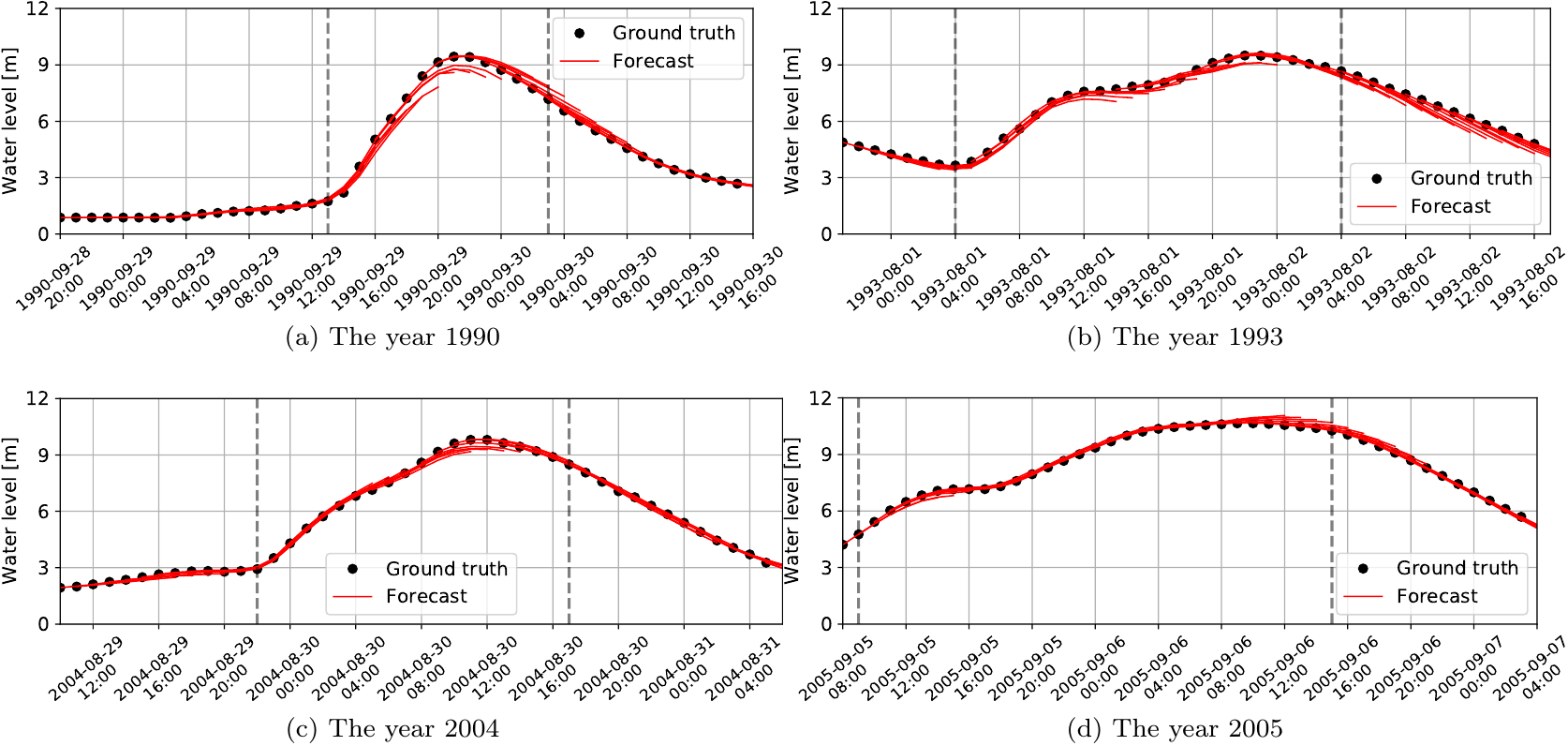}
    \caption{Forecast results for the Hiwatashi gauging station in (a) 1990, (b) 1993, (c) 2004, and (d) 2005. Each event relates to the maximum water rise event of the corresponding year, and the events are arranged in ascending order of the maximum water level; namely, panel (d) shows the values of maximum water rise for all events. The black points denote the observed river stages, and the red solid lines show the forecasts from the observed river stage up to 6 h ahead. The gray dashed lines show the evaluation intervals of the RMSEs.}
    \label{fig:hiwatashi_series}
\end{figure}

\begin{figure}
    \includegraphics[width=1.\linewidth]{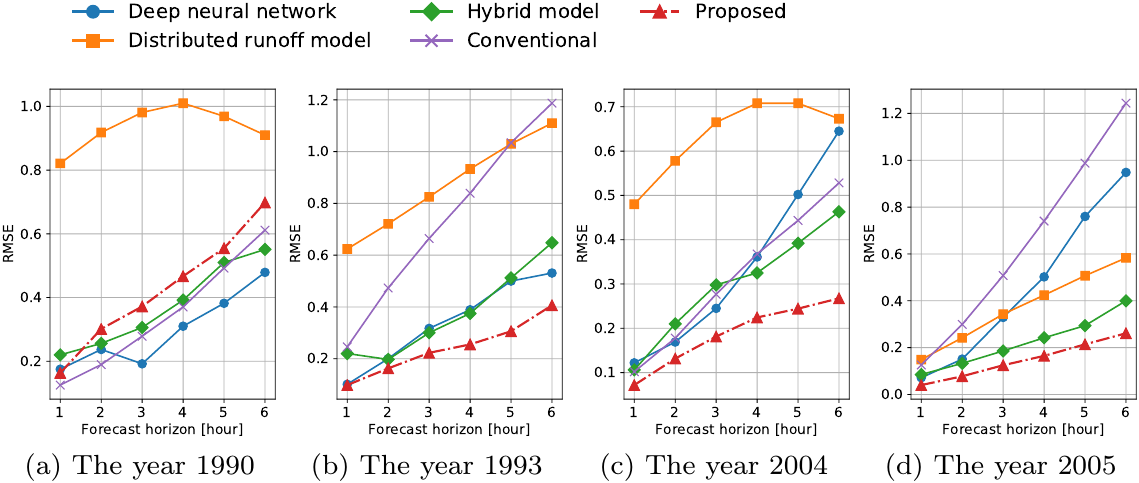}
    \caption{RMSEs for the Hiwatashi gauging station in (a) 1990, (b) 1993, (c) 2004, and (d) 2005. We compared the proposed forecast approach with three existing methods: the deep neural network, the distributed runoff model, and the hybrid model of the deep neural network and distributed runoff model. Note that the existing results are scanned values from Figure 10 in \citeA{hitokoto2017}.}
    \label{fig:hiwatashi_rmse}
\end{figure}

The proposed method accurately forecasted for all test cases, including those of the maximum water levels (Figure~\ref{fig:hiwatashi_series}).
We also compared the performance with the other methods described by \citeA{hitokoto2017} using the root mean squared error (RMSE).
Although the distributed runoff model and the hybrid model require many kinds of hydrological details, the proposed method---which requires information only on the series of the rainfall and river stage---achieved the best RMSE result in three cases out of four (Figure~\ref{fig:hiwatashi_rmse}).
The higher the maximum water level, the better the accuracy of the proposed method compared to the other methods, and Figures~\ref{fig:hiwatashi_rmse}(b), (c), and (d) show that the best performance is obtained by the proposed method.
In the case of Figure~\ref{fig:hiwatashi_rmse}(a), although the proposed method did not achieve the best performance, the method forecasted with good accuracy and provided useful information for decision makers (Figure~\ref{fig:hiwatashi_series}).
This is because the behavior of water rise is the simplest and the magnitude of water level is the smallest among all the cases.
Therefore, it is easy to forecast the river stage and the result shows that the all the methods, except the distributed runoff model, performed equally well and competitively in this case.

\subsection{Kagetsu Gauging Station on the Oyodo River System}
\begin{figure}
    \centering
    \includegraphics[width=.8\linewidth]{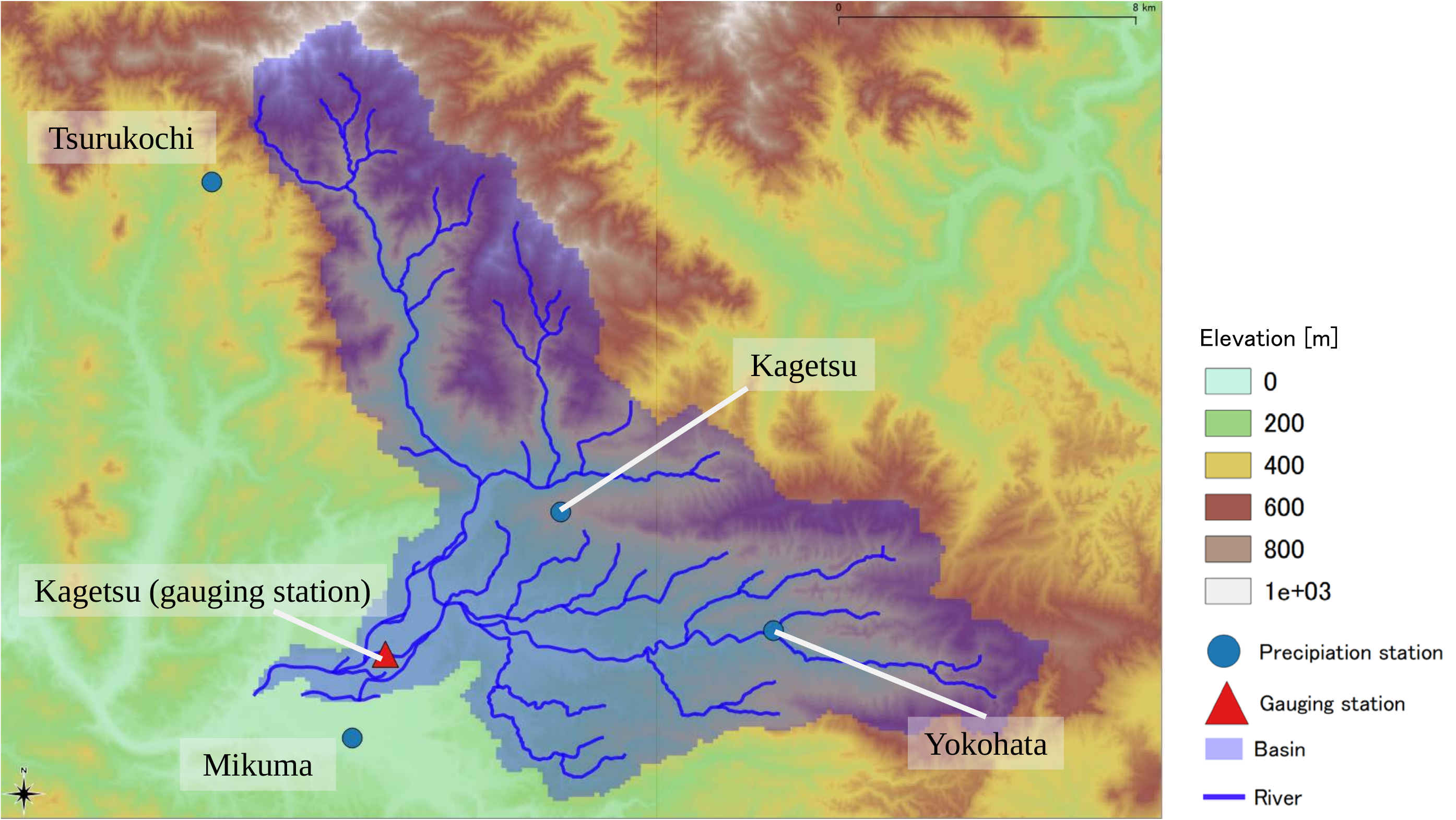}
    \caption{Locations of the Kagetsu gauging station, related rivers, and nearby stations. The figure is created using the river and basin data from the National Land Numerical Information and the adjusted elevation data created by \citeA{yamazaki2018}.}
    \label{fig:kagetsu_basin}
\end{figure}

\begin{figure}
    \centering
    \includegraphics[width=.5\linewidth]{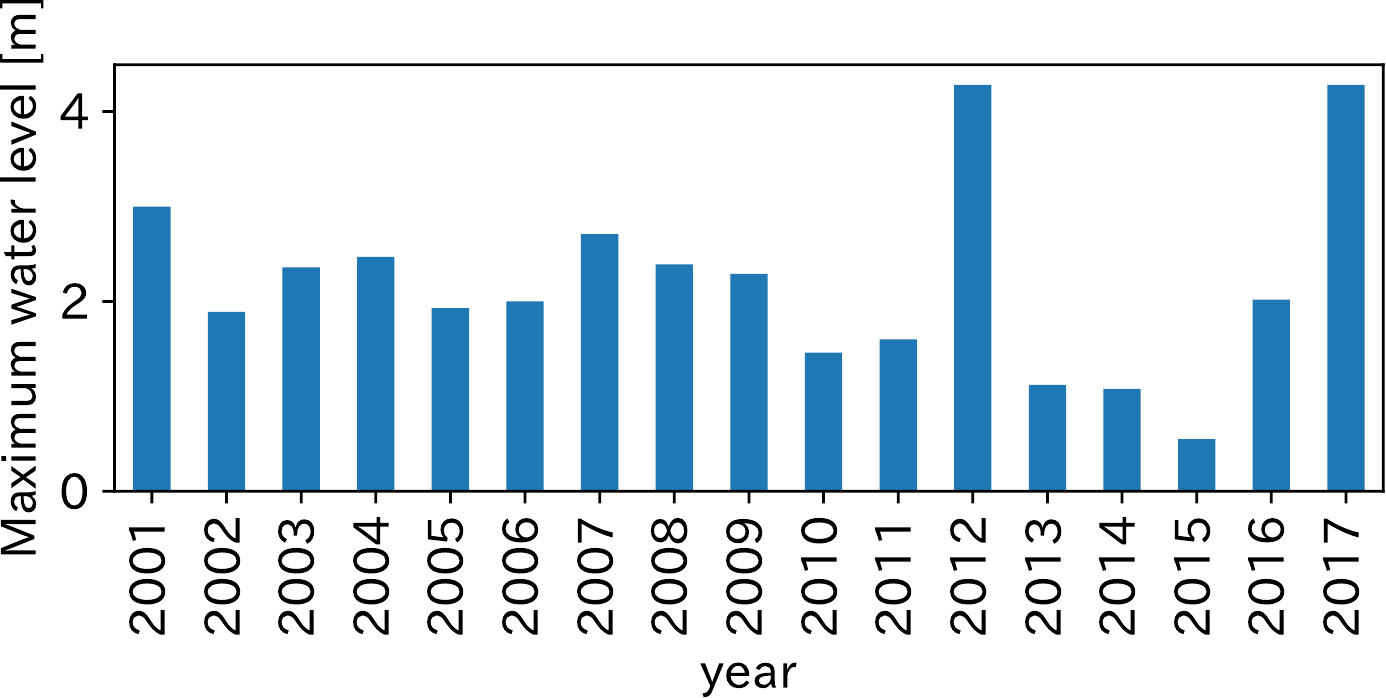}
    \caption{Maximum water levels for each year for the Kagetsu gauging station. We evaluated the water rise events for 2012 and 2017.}
    \label{fig:kagetsu_max_heights}
\end{figure}

We demonstrated the practicality of the proposed method using data of the Kagetsu gauging station on the Chikugo river system for the 2012 and 2017 heavy rainfall events in the northern Kyushuthe.
Namely, we evaluated the lead time for evacuation, accuracy of unexperienced water levels, and sensitivity analysis of water rise events for actual usage.
We sourced the past hourly data on water levels from the Kagetsu gauging station and four precipitation stations (Tsurukochi, Kagetsu, Yokohata, and Mikuma) from 2001 through 2017 from the website of the Water Information System \cite{mizumizuweben}.
The locations of the river, gauging stations, and precipitation stations appear in Figure~\ref{fig:kagetsu_basin}.
We picked 36 water rise events, wherein one event includes samples from 36 h before the time when the level first exceeded 0.9 m, to 24 h after the time when it last exceeded this level.
Note that the local government specifies the evacuation warning water level as 2.2 m, and 0.9 m is the lowest warning water level until the end of 2017 heavy rainfall events in the northern Kyushu.
We estimated the 6-hours-ahead forecast for the 2012 and 2017 heavy rainfall events in the northern Kyushu, both of which recorded the highest water levels at the corresponding times (Figure~\ref{fig:kagetsu_max_heights}).
In these cases, to assume as realistic a situation as possible, we selected embeddings using only the data up to 2011 in advance, and we forecasted the test cases (2012 and 2017) using the samples up to the corresponding time.

\begin{figure}
    \centering
    \includegraphics[width=.6\linewidth]{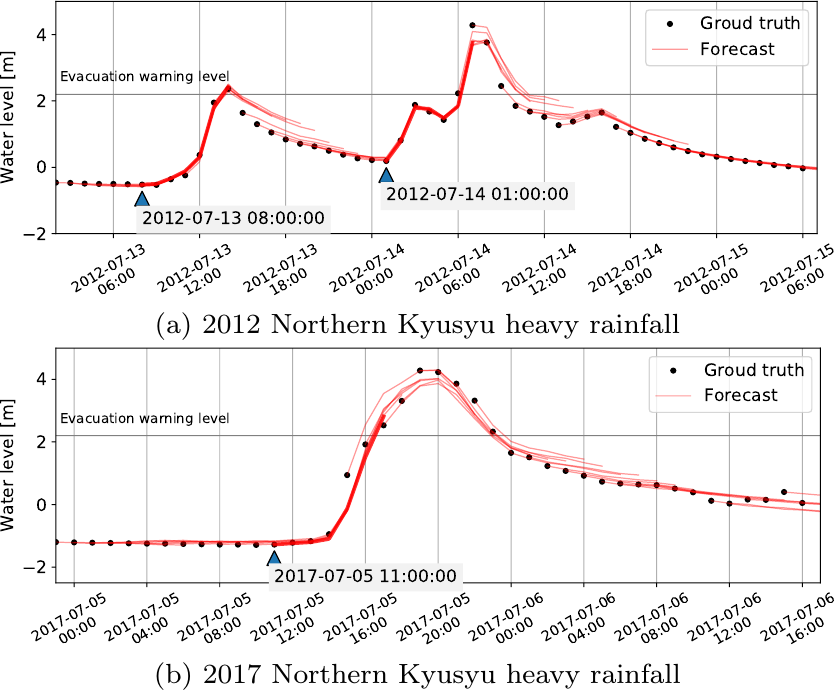}
    \caption{Forecast series for the Kagetsu gauging station. The black points show the observed water levels, and the red thin lines denote the corresponding up-to-6-hours-ahead forecasts. The bold red lines indicate the forecast result when the excesses of the evacuation warning levels were forecasted, and the blue triangle markers indicate the time when the excesses were forecasted.}
    \label{fig:kagetsu_series}
\end{figure}

The proposed method provided an appropriate forecast (Figure~\ref{fig:kagetsu_series}) although the water level radically rose in a few hours and both events caused previously unexperienced flood magnitudes (Figure~\ref{fig:kagetsu_max_heights}).

We also evaluated the lead time, which is defined as the interval from the time when the proposed method forecasted the exceedance of the evacuation warning level (2.2 m) to the actual exceedance time.
The proposed method forecasted the exceedance of the evacuation warning level 6 h earlier for most cases (Table~\ref{tab:lead_time}). Thus, the proposed method was able to provide useful information for decision makers under these severe conditions.

\begin{table}
    \centering
    \caption{Lead time for evacuation. We define the lead time as the difference between the time the method forecasted the exceedance in the evacuation warning level and the actual time of the exceedance.}
    \label{tab:lead_time}
    \begin{tabular}{c|p{2cm}|p{2.5cm}|p{3.cm}|c} \hline
        Date       & Actual exceeded time & Forecasted exceeded time & Time when excess was forecasted & Lead time \\ \hline
        2012-07-13 & 14:00                & 14:00                    & 08:00                           & 6hours    \\ \hline
        2012-07-14 & 06:00                & 07:00                    & 01:00                           & 5hours    \\ \hline
        2017-07-05 & 17:00                & 16:00                    & 10:00                           & 7hours    \\ \hline
    \end{tabular}
\end{table}

\begin{figure}
    \centering
    \includegraphics[width=.6\linewidth]{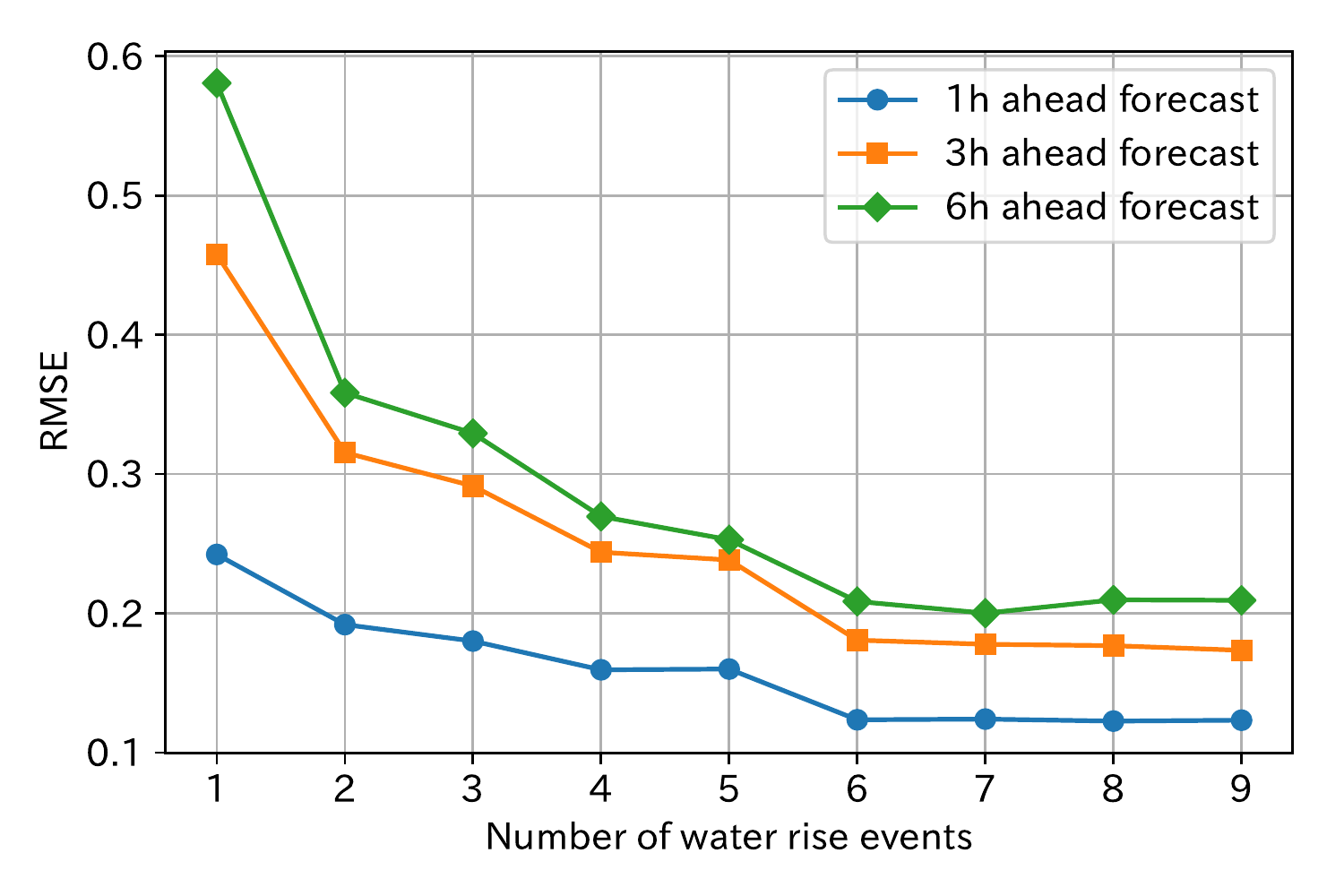}
    \caption{RMSE by the number of water rise events at the Kagetsu gauging station. The blue line, orange line, and green line show the RMSEs of the 1-hour-ahead, 3-hours-ahead, and 6-hours-ahead forecasts, respectively.}
    \label{fig:kagetsu_rmse}
\end{figure}

\begin{figure}
    \centering
    \includegraphics[width=.6\linewidth]{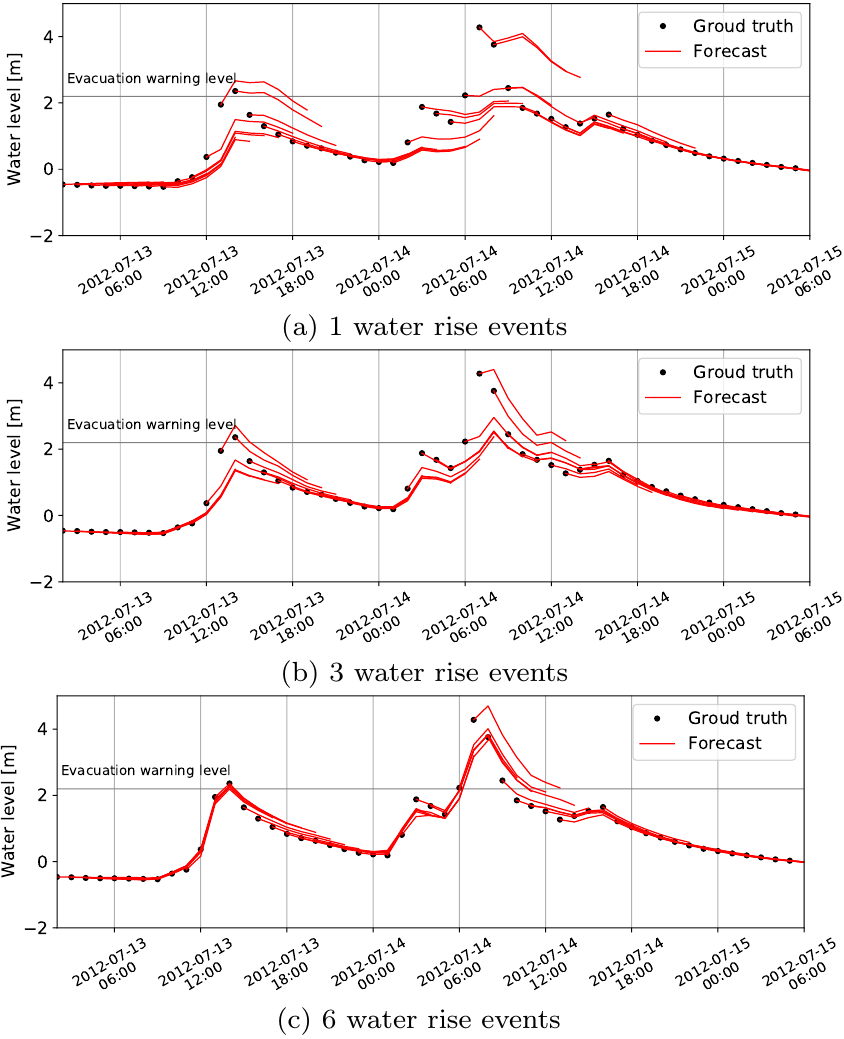}
    \caption{Forecast results for the Kagetsu gauging station for the 2012 northern Kyushu heavy rainfall event. Panels (a), (b), and (c) show the forecast result using one water rise event, three water rise events, and six water rise events, respectively.
    }
    \label{fig:kagetsu_sensitivity_series}
\end{figure}

We also conducted a sensitivity analysis on the number of water rise events and the forecast accuracy.
We evaluated the forecast error with seven water rise events---which exceeded 0.9 m from 2011 through 2012, including the 2012 heavy rainfall event in the northern Kyushu---by changing the number of water rise events.
Although the method could not provide a correct forecast if information on only one water rise event was provided (Figure~\ref{fig:kagetsu_sensitivity_series}(a)), the accuracy was significantly improved as the number of events increased (Figure~\ref{fig:kagetsu_rmse}).
Information on three water rise events enabled the method to forecast the water rise trend, and information on six events enabled the method to forecast the previously unexperienced water rise from 06:00 through 07:00 on July 14, 2012 (Figures~\ref{fig:kagetsu_sensitivity_series}(b) and (c)).

\section{Discussion and Conclusion}
The proposed method achieved the best overall performance for the Hiwatashi gauging station compared with other representative methods, including data-driven and hydrological approaches.
Comparing with existing methods, the proposed method accurately forecasted the unexperienced water level (Figures~\ref{fig:hiwatashi_rmse}(d) and \ref{fig:hiwatashi_series}(d)), and this result can be attributed to the correction term of equation~(\ref{eq:forecast}).
We also validated this property using a flood competition dataset, and the proposed method achieved the best performance for the test data, including the highest peak water level.
See detail in \ref{sec:annex}.

Besides the advantage of accuracy, the method offers several improvements over conventional approaches in terms of practical use.
The first concerns the maintainability of the system.
Once the suitable embeddings are selected, we can simply append the latest observations to $\mathcal{T}_{train}$ and can improve forecast performance without re-training.
Note that this is a common property of local model approaches based on delay embedding.
The second advantage is the stability of the forecast.
In contrast to black box function approaches, the proposed map interpolates or extrapolates only on the basis of the reconstructed trajectories.
Therefore, the method does not yield anomalous forecasts even if the query is out of the range of the training data, as shown in the examples.
Simultaneously, we can easily analyze the cause of the anomaly forecast even if it occurs by simply verifying the neighboring trajectories.
Another advantage is that the method requires a small amount of data in fewer kinds compared to previously used approaches.
The method does not require any other hydrological information besides that on rainfall and river stage series.
In addition, the method can be implemented using only a small amount of data, as shown in the sensitivity analysis.
With regard to the example of the Kagetsu gauging station, we could forecast the river stages using only six water rise events.
This is because the multiple embeddings approaches were originally proposed for short time series.
The last advantage relates to the low cost of parameter tuning.
The proposed map does not require any parameter tuning because all the parameters are automatically determined by the optimizations; thus, we can easily apply it to any site as long as the river stage and rainfall data are available.

In this study, we tested the proposed method for relatively simple circumstances.
We need to evaluate the method under more complex situations, such as rivers influenced by dam discharge or tides.
In addition, we need to consider errors in the rainfall forecasts, which can introduce significant issues in practical use.
We plan to tackle these problems in a future study.

\appendix
\section{Delay Embedding} \label{sec:embedding}
\begin{figure}
    \centering
    \includegraphics[width=.7\linewidth]{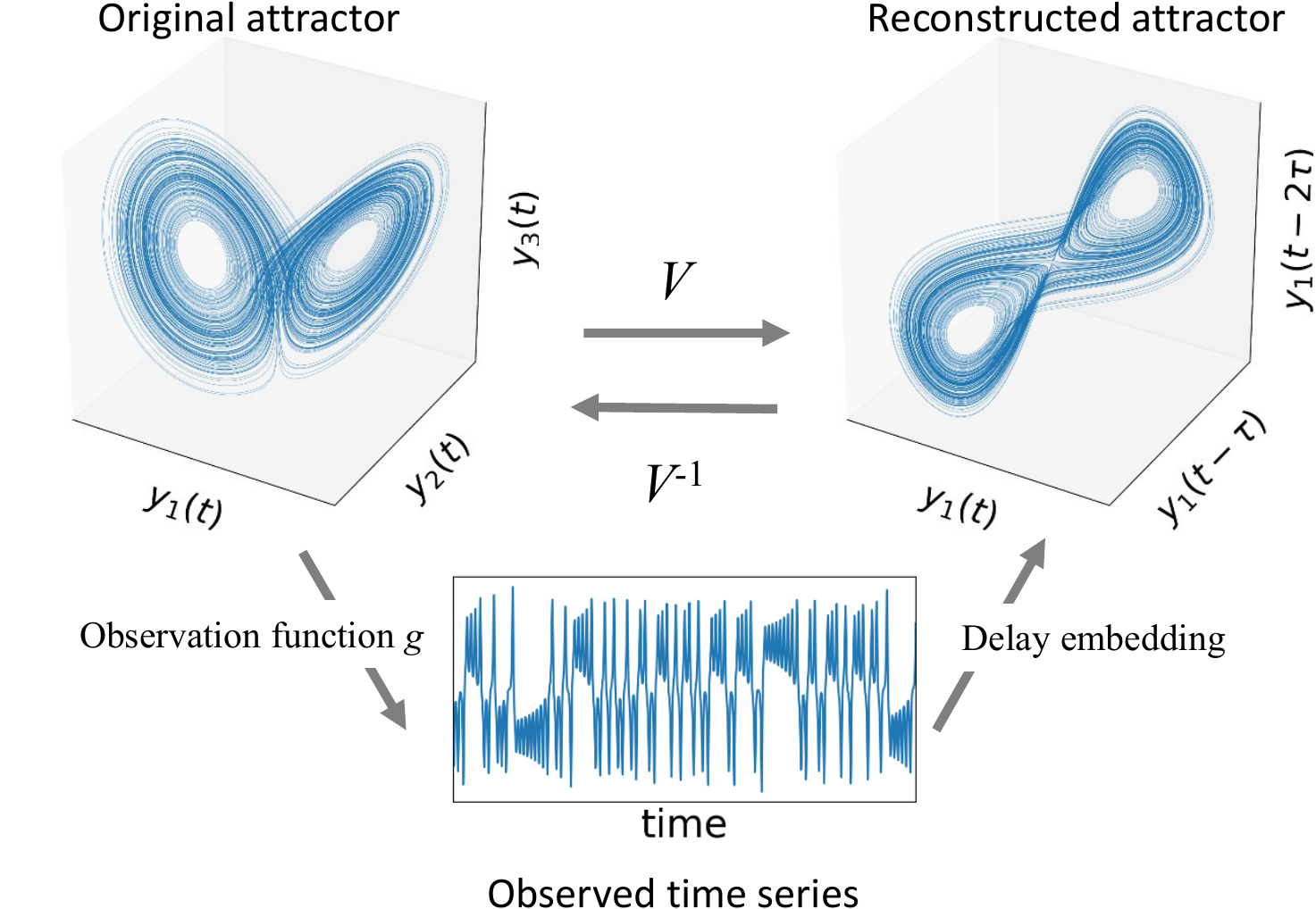}
    \caption{Schematic of delay embedding. We can reconstruct the original attractor from the observed time series. Note that the map from the original attractor to the reconstructed attractor has a one-to-one correspondence.}
    \label{fig:embedding}
\end{figure}

One of the most common approaches for reconstructing underlying dynamics is delay embedding.
When we observe only $y_1 (t) \in \mathbb{R}$ from equation~(\ref{eq:system}), we consider the following map $V:\mathbb{R}^m \to \mathbb{R}^E$ given by
\begin{equation}
    V(x)=[g(x), g(f(x)), g(f^2(x), ..., g(f^{E-1}(x))].
\end{equation}
According to the embedding theorems \cite{Takens1981,Sauer1991a}, the sufficient condition for the embedding of map $V$ is $E> 2m$.
In particular, with the observed time series ${y_1(t)}$, a lag $\tau$, and an appropriate embedding dimension $E$, the following delay coordinates $v(t) \in \mathbb{R}^E$ reconstruct the original attractor:
\begin{equation}
    v(t)=[y_1 (t), y_1 (t-\tau), ..., y_1 (t-(E-1)\tau)].
\end{equation}
See the schematic in Figure~{\ref{fig:embedding}} for details.

These embedding theorems have been extended to external forcing \cite{Stark1999}, as well as multivariate data and nonuniform embeddings \cite{Deyle2011}.
For the multivariate data and nonuniform embeddings, we can reconstruct the original attractor with $y(t)=[y_1 (t),y_2 (t),...,y_n (t)]$, $\sigma_i \in \{1,2,...,n\}$, and lags $\tau_i \in \{0,1,...,l-1\}$, using the following delay coordinates:
\begin{equation} \label{eq:delay_embedding}
    v(t)=\left[ y_{\sigma_1}(t-\tau_1), y_{\sigma_2}(t-\tau_2),...,y_{\sigma_E}(t-\tau_E) \right],
\end{equation}
where $\tau_j=0 \ \exists j \in \{1,2,...,n\}$, and no duplication is allowed for any element.

\section{Forecast Exploiting Suboptimal Embeddings} \label{sec:ga}
Here, we define the time indices of training data as $\mathcal{T}_{train}=\{t \mid t < 0\}$.
We split $\mathcal{T}_{train}$ into $K$, and define the $K'$-th split data as $\mathcal{T}_{train}^{K'}$.
In this study, we split the data by every water rise event for our Japanese river data,
and we split every single year's data into two for the flood competition dataset.
We solve the following optimization problem to obtain suboptimal embeddings \cite{Okuno2019a}:
\begin{equation} \label{eq:fitness}
    {\rm minimize}_{e \in \mathcal{E}} \ \sum_{t \in \mathcal{T}_{train}^{K'}} \sum_p \|\hat{y}_r^e (t+p|t) - y_r (t+p)\|,
\end{equation}
where $\|\cdot\|$ is an appropriate norm, and $\hat{y}_r^e (t+p|t)$ is the $p$-steps-ahead forecast at time $t$ based on embedding $e$. We employ the $L^2$ norm in this paper.
We first solve equation~(\ref{eq:fitness}) using a genetic algorithm, storing all the evaluated solutions in the process of optimization.
Then, we select the top $M$ embeddings that satisfy the following condition:
\begin{equation} \label{eq:embedding_condition}
    d(e_i, e_j) \geq \theta, \ \forall i,j; i \neq j,
\end{equation}
where $d(\cdot,\cdot)$ is the Hamming distance.
We set $\theta=3$ and $M=3$ in this study.
Using the suboptimal embeddings, we compute $\hat{k}_p$ to minimize the combined in-sample forecast error as follows:
\begin{eqnarray}
    Y_k(t+p|t) := 1/k \sum_{i=1}^k \hat{y}_r^{\mathcal{I}_p(i)} (t+p|t),  \\
    \hat{k}_p = {\rm argmin}_{k=1,2,...,P} \sum_{t \in \mathcal{T}_{train}} [Y_k(t+p|t) - y_r(t)]^2, \\
    \hat{y}_r(t+p|t) = Y_{\hat{k}_p}(t+p|t),
\end{eqnarray}
where $\mathcal{I}_p$ is a tuple of forecast indices sorted by the in-sample error corresponding to the whole $\mathcal{T}_{train}$.
Forecast $\hat{y}_r (t+p|t)$ is computed as follows:
\begin{equation}
    \hat{y}_r (t+p|t) =  \sum_{k=1}^{\hat{k}_p} \hat{y}_r^{\mathcal{I}_p(k)}(t+p|t) \  / \ \hat{k}_p.
\end{equation}

\section{Application to Flood Forecasting Dataset} \label{sec:annex}
We also tested the proposed method with a flood forecasting dataset named ``Artificial Neural Network Experiment (ANNEX 2005/2006)'' \cite{Dawson2005a}.
The dataset contains water level series at four locations---a target site and three upstream sites---and precipitation series at five locations.
The dataset contains six hourly samples for three periods: January 10, 1993 to March 31, 1994 (728 samples) and January 10, 1995 to March 31, 1996 (732 samples) (for training), and January 10, 1994 to March 31, 1995 (728 samples) (for testing).
Note that the peak water levels in the training data are 4.122 m and 4.997 m, and the peak level in the test data is recorded as 5.746 m.
The catchment covers an area of $3315\textrm{km}^2$.

We forecasted the water level of the target site for 6, 12, 18, and 24 h using our proposed method.
We also forecasted the water level using a conventional local linear model \cite{Farmer1987} based on delay embedding and compared the findings with existing machine learning results \cite{Okuno2019}, namely, a recurrent neural network with long short-term memory (LSTM) \cite{Hochreiter:1997:LSM:1246443.1246450}, support vector regression (SVR) \cite{Boser:1992:TAO:130385.130401}, and random forest regression \cite{Breiman2001}.

\begin{figure}
    \centering
    \includegraphics[width=0.9\linewidth]{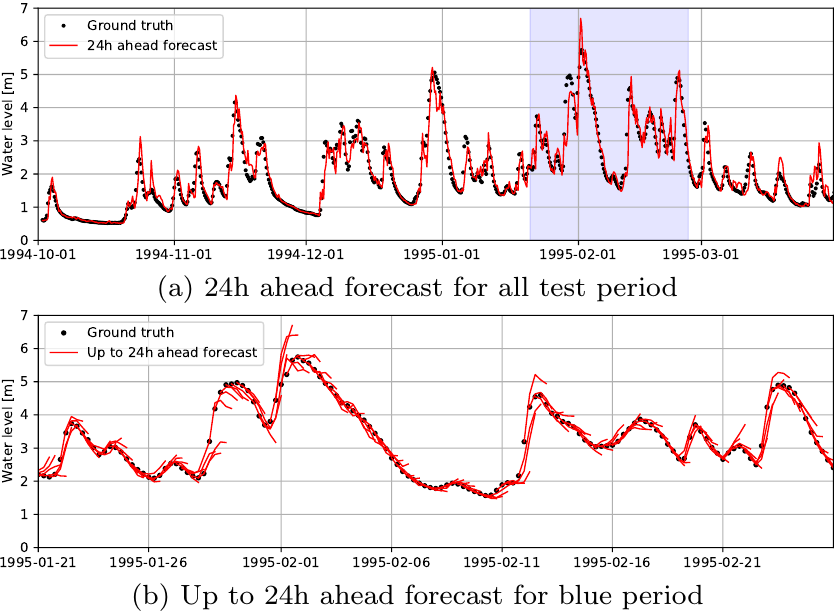}
    \caption{Forecast results for the ANNEX 2005/2006 dataset. Panel (a) shows a comparison of the 24-hours-ahead forecast (the red line) and the ground truth (the black dots) for all test data. Panel (b) shows a comparison of up to 24-hours-ahead forecasts (the red lines) and the ground truth (the black dots) for the blue period in Panel (a), which includes the highest water level for all periods. Each red line represents the forecasts (6 h, 12 h, 18 h, and 24 h) of the corresponding black dot.}
    \label{fig:annex}
\end{figure}

The proposed forecast yielded the best performance for 12- to 24-hours-ahead forecasts in contrast to the conventional delay embedding method, which provided the worst result for the 24-hours-ahead forecast.
Although the accuracy of the 6-hours-ahead forecast was the second best, the difference was negligible.
As shown in Table~\ref{tab:flood} and Figure~\ref{fig:annex}, the proposed method forecasted the river stage with good overall accuracy.
Although the test data include the unexperienced peak water level, the method forecasted the peak without underestimation due to the correction term in equation~(\ref{eq:forecast}).

\begin{table}
    \centering
    \begin{tabular}{c|c|c|c|c|c|c}
        \hline
        Steps [h] & Proposed       & Conventional & LSTM  & SVR            & Random Forest \\ \hline
        6         & 0.074          & 0.086        & 0.123 & \textbf{0.073} & 0.079         \\ \hline
        12        & \textbf{0.153} & 0.177        & 0.213 & 0.177          & 0.185         \\ \hline
        18        & \textbf{0.223} & 0.264        & 0.291 & 0.268          & 0.275         \\ \hline
        24        & \textbf{0.317} & 0.378        & 0.372 & 0.365          & 0.363         \\ \hline
    \end{tabular}
    \caption{\label{tab:flood}RMSEs of the ANNEX 2005/2006 flood dataset computed using the proposed method (Proposed), a conventional local linear method (Conventional), and existing results by machine learning methods \cite{Okuno2019}. The best score for each time step is indicated in boldface.}
\end{table}

%
%
%
%
%
%
%
\acknowledgments
We thank Prof. Christian W. Dawson for the ANNEX 2005/2006 dataset.
Interested readers are requested to contact Prof. Dawson directly to obtain the dataset.
We thank the Ministry of Land, Infrastructure, Transport and Tourism for providing access to the web service of the Water Information System.
The data for the Kagetsu and Hiwatashi gauging stations can be downloaded directly from the website.
We thank Dr. Yoshito Hirata for introducing us to the ANNEX 2005/2006 dataset and for suggestions and feedback on the early stage of this study.
This research was partially supported by Kozo Keikaku Engineering Inc., JSPS KAKENHI Grant Number JP15H05707, and World Premier International Research Center Initiative, Ministry of Education, Culture, Sports, Science and Technology, Japan.

%
%

%
%
%
%
%

\bibliography{library}
\end{document}